\documentclass[journal]{IEEEtran}
\IEEEoverridecommandlockouts
\usepackage{etoolbox}
\usepackage{cite}
\usepackage{amsmath,amssymb,amsfonts}
\usepackage{algorithmic}
\usepackage{graphicx}
\usepackage{multirow}
\usepackage[flushleft]{threeparttable}
\usepackage{textcomp}
\usepackage{array}
\usepackage[normalem]{ulem}
\newcolumntype{P}[1]{>{\centering\arraybackslash}m{#1}}
\begin{document}
\title{Stateful Logic using Phase Change Memory}
\author{Barak Hoffer, Nicolás Wainstein, \IEEEmembership{Member, IEEE}, Christopher M. Neumann, Eric Pop, \IEEEmembership{Fellow, IEEE}, Eilam Yalon, \IEEEmembership{Member, IEEE}, and Shahar Kvatinsky, \IEEEmembership{Senior Member, IEEE}
\thanks{Manuscript received XXX XX, 2022; revised XXX XX, 2022;
accepted XXX XX, 2022. Date of publication XXX XX, 2022; date
of current version XXX XX, 2022. This work was partially supported by the European Research Council through the European Union's Horizon 2020 Research and Innovation Programe under Grant 757259, and partially supported by the European Research Council through the European Union's Horizon Europe Research and Innovation Programe under Grant 101069336.}
\thanks{B. Hoffer, N. Wainstein, Eilam Yalon and S. Kvatinsky are with
the Andrew and Erna Viterbi Faculty of Electrical and Computer Engineering, Technion-Israel Institute of Technology, Haifa 3200003, Israel (email: shahar@ee.technion.ac.il).}
\thanks{C. M. Neumann and E. Pop are with the Department of Electrical Engineering, Stanford University, Stanford, CA 94305 USA.}}

\markboth{IEEE Journal on, VOL. XX, NO. XX, XXXX 2022}
{Hoffer \MakeLowercase{\textit{et al.}}: Experimental Demonstration of Stateful Logic using Phase Change Memory}

\maketitle
\IEEEpubid{\begin{minipage}{\textwidth}\ \\[12pt] \centering \copyright 2022 IEEE. Personal use of this material is permitted. Permission from IEEE must be obtained for all other uses, in any current or future media, including reprinting/republishing this material for advertising or promotional purposes, creating new collective works, for resale or redistribution to servers or lists, or reuse of any copyrighted component of this work in other works.\end{minipage}}

\begin{abstract}
Stateful logic is a digital processing-in-memory technique that could address von Neumann memory bottleneck challenges while maintaining backward compatibility with standard von Neumann architectures. 
In stateful logic, memory cells are used to perform the logic operations without reading or moving any data outside the memory array.
Stateful logic has been previously demonstrated using several resistive memory types, mostly by resistive RAM (RRAM). Here we present a new method to design stateful logic using a different resistive memory - phase change memory (PCM). We propose and experimentally demonstrate four logic gate types (NOR, IMPLY, OR, NIMP) using commonly used PCM materials. Our stateful logic circuits are different than previously proposed circuits due to the different switching mechanism and functionality of PCM compared to RRAM. Since the proposed stateful logic form a functionally complete set, these gates enable sequential execution of any logic function within the memory, paving the way to PCM-based digital processing-in-memory systems.
\end{abstract}

\begin{IEEEkeywords}
phase-change-memory (PCM), processing-in-memory (PIM), stateful-logic
\end{IEEEkeywords}

\section{Introduction}
\IEEEpubidadjcol
\label{sec:introduction}
\IEEEPARstart{F}{or} the last 75 years, computers have been typically designed in the von Neumann architecture, which separates the memory from the processing units.
While their programming model is simple, incessant data movement limits system performance because memory access time is often substantially longer than the computing time. This bottleneck has worsened over the years since CPU speed has improved more than memory speed and bandwidth (the so-called `memory wall')~\cite{Hennessy2017}.
One attractive approach to deal with this problem is processing-in-memory (PIM), which suggests adding computation capabilities to the memory. PIM reduces the need for costly (in terms of processing-speed, bandwidth, and energy) chip-to-chip transfers, thus yielding higher performance and energy efficiency~\cite{Mutlu2020}.

An increasing number of applications from high-performance computing (HPC) to databases, data analytics and deep neural networks require higher memory capacity to meet the needs of workloads with large data sets. DRAM scaling has slowed down in the last years, and it has become a Herculean task to improve its capabilities further~\cite{Mandelman2012,Shiratake2020}. Thus, new technologies, such as resistive random-access memory (RRAM), conductive-bridge RAM (CBRAM), and phase-change memory (PCM), are being explored~\cite{Yu2016}. These technologies offer increased memory capacity and add non-volatility, enabling persistent memory. These types of persistent memories are usually referred to as Storage Class Memory (SCM)~\cite{Fong2017,Chen2020}, as they combine both storage and memory characteristics.
With SCM, applications stand to benefit from the availability of large-capacity memory, but the performance will still be limited by the incessant data movement between the CPU and memory.

\textit{Stateful logic}~\cite{Borghetti2010,Kvatinsky2014} is a processing-in-memory technique based on memristive memory technologies, \textit{e.g.}, RRAM or CBRAM.
In stateful logic gates, the input and output are represented in the form of resistance, and the result is written during the computation directly to the output memory cell without reading the input cells beforehand or moving any data outside the memory array~\cite{Reuben2017}.
When the stateful logic gates are functionally complete (\textit{e.g.}, NOR gates), any desired function can be computed using a sequence of stateful logic operations~\cite{BenHur2020,Ronen2022,Leitersdorf2022}.
Stateful logic enables PIM architectures such as the \textit{memristive memory processing unit} (mMPU)~\cite{Haj-Ali2018b} and RACER~\cite{Truong2021} that offer massive intrinsic parallelism, high-performance, and energy-efficient processing, while maintaining backward compatibility with von Neumann architectures.
\IEEEpubidadjcol
Prior studies on stateful logic mostly focused on bipolar RRAM devices~\cite{Jang2017,Bae2017,Jang2018,Zhong2018,Shen2019,Kim2019,Nuo2020,Kim2020,Hoffer2020,Kim2021,Liu2021}, which still suffer from reliability and variability issues~\cite{Chen2020,Hyun2020} and are unavailable commercially in large scale.
PCM is a more mature resistive technology that offers fast operation speed, low power, good reliability, and high density integration, while being already used commercially~\cite{Cheng2015,Fong2017,Cheng2020,Gong2020}, $e.g.$, in the Intel Optane technology~\cite{Choe2017}. However, previous studies of computation using PCM mainly focused on analog neuromorphic computation~\cite{Suri2011,Burr2015,Sebastian2017,Boybat2018,Joshi2020,Nandakumar2020,Sebastian2019,Xu2020} or non-stateful binary logic operations~\cite{Li2013,Lu2016,Loke2014,Giannopoulos2020}. Since the \emph{switching mechanism of PCM is unipolar and completely different than RRAM}, to achieve stateful logic using PCM devices, new circuit topologies and voltage schemes are needed. To the best of our knowledge, only a single stateful logic method previously proposed for PCM, which was mentioned and demonstrated for a single test cycle in ~\cite{Cassinerio2013}. In this method, called input-output transfer, three sequential voltage pulses are required to perform a single AND operation.

In this paper, we present and experimentally demonstrate a new method to perform stateful logic operations using PCM in a single step. We demonstrate four different logic gates (NOR, IMPLY, OR, and NIMP) with robust and repeatable results.
Our logical set is functionally complete, enabling sequential execution of any logic function in-memory. The gates are compatible with the memory crossbar structure and can be applied in parallel on multiple rows~\cite{Haj-Ali2018b}.

\section{Stateful logic with Phase-change memory}
\begin{figure}[t]
\includegraphics[width=\linewidth]{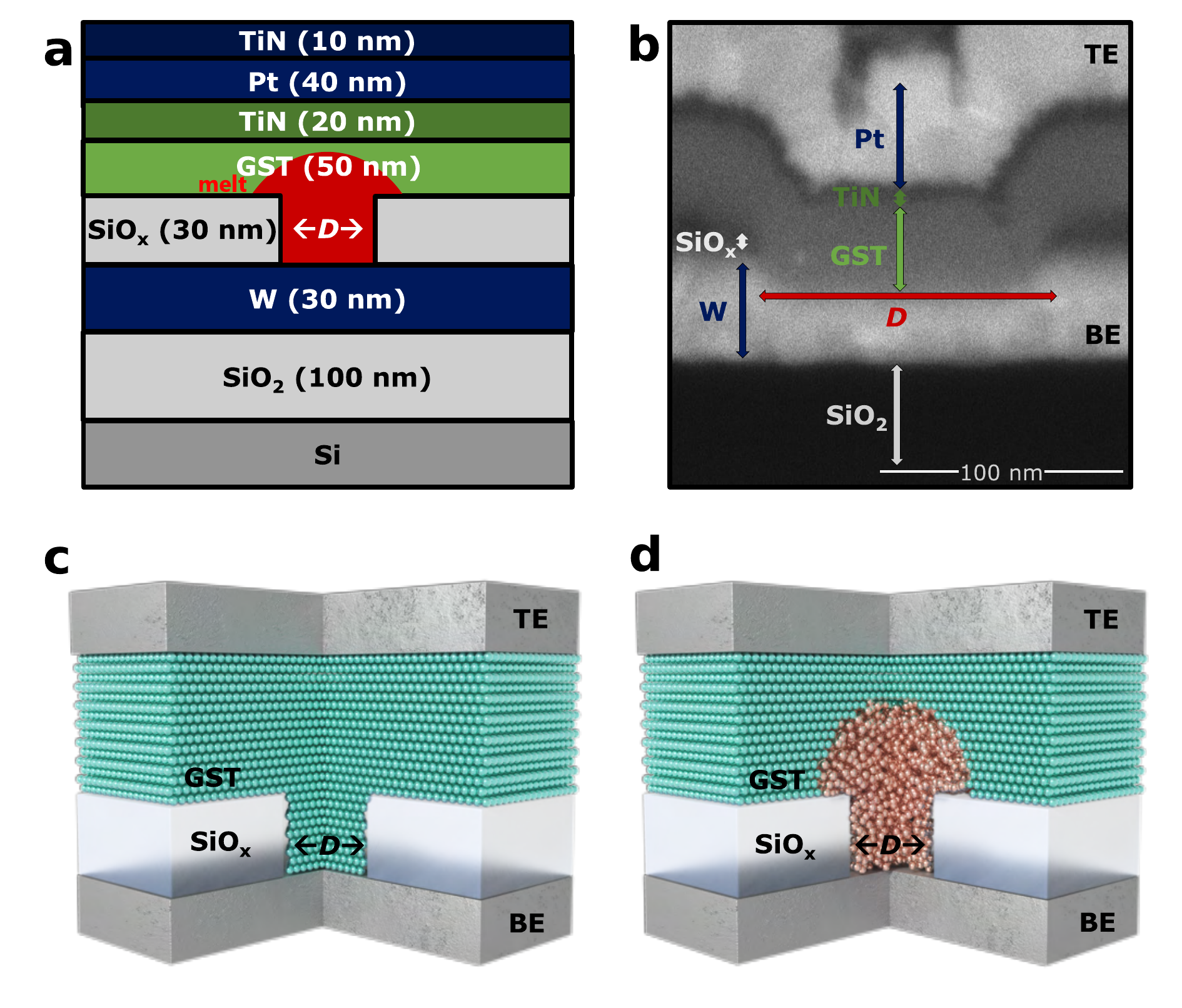}
\caption{Confined PCM cell used in this work. (a) Cross-section schematic, (b) Cross-section scanning electron microscopy (SEM), (c)-(d) 3D cartoon of the crystalline and amorphous states. Evaporation and etching of tungsten (W) is used to form the bottom electrode (BE). Sputtering and liftoff are used to pattern the Ge\textsubscript{2}Sb\textsubscript{2}Te\textsubscript{5} (GST) layer, the TiN/Pt top electrode (TE) and contact pads. GST is patterned into a confined area with diameter $D \sim$ 125 nm.}
\label{fig:device}
\end{figure}
\begin{figure}[t]
\includegraphics[width=\linewidth]{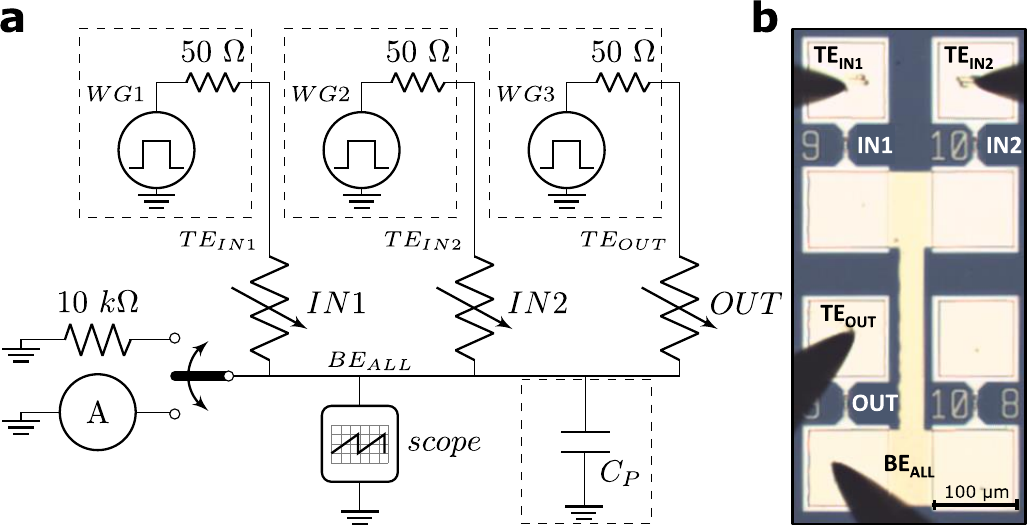}
\caption{Experimental setup. (a) Schematic and (b) optical top-view. Three cells are connected to a shared bottom electrode. Three waveform generators are connected to the top electrode of each cell. The bottom electrode can be switched between: floating, grounded, and grounded using a 10~k$\Omega$ resistor. $C_P$ marks the total parasitic capacitance at the shared node caused by the pads and probes.}
\label{fig:setup}
\end{figure}

\begin{figure*}[t]
\centering
\includegraphics[width=\linewidth]{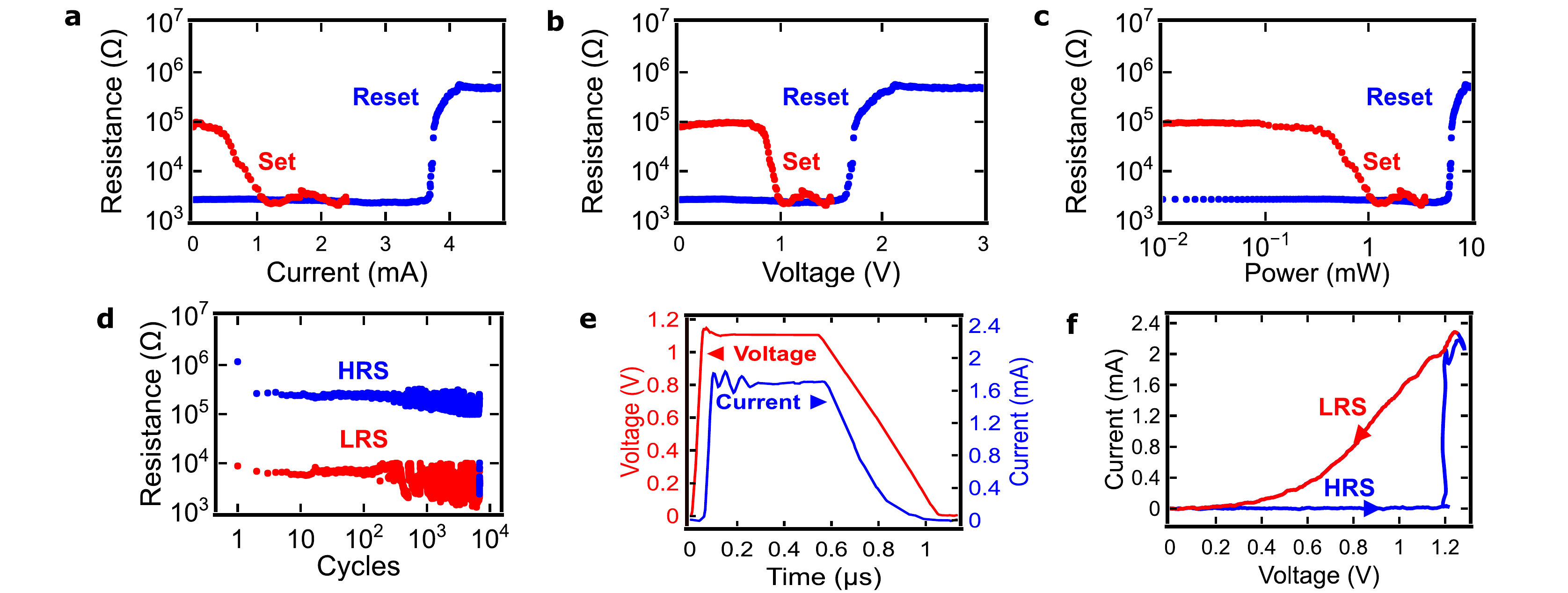}
\caption{In-house fabricated PCM device characteristics. DC read resistance during set and reset versus vs (a) programming current, (b) applied voltage, and (c) power. (d) Endurance data for a representative PCM device. The device was cycled using a write-verify scheme. The voltages for set and reset are 1.2~V and 3.0~V, respectively. The rise/width/fall of the pulses for set and reset are 30/500/500~ns and 30/50/30~ns, respectively.
(e) Current and voltage across the PCM cell during set operation. Threshold switching occurs within $\sim$100~ns and the voltage is kept high for the remaining time to complete the crystallization process.
(f) I-V transition of the device from the amorphous state (HRS) to the crystalline state (LRS) showing its threshold switching voltage, 1.2V. The I-V was measured using a triangular voltage ramp with 1~$\mu$s rise and fall times.}
\label{fig:characterize}
\end{figure*}
\begin{figure*}[t]
\includegraphics[width=\linewidth]{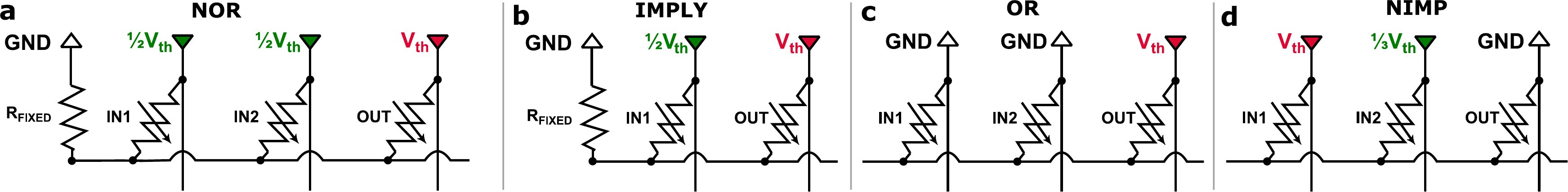}
\caption{Schematics for PCM stateful logic gates mapping to the crossbar structure. (a) NOR gate. (b) IMPLY gate. (c) OR gate. (d) NIMP gate.}
\label{fig:crossbar}
\end{figure*}

\subsection{Phase-Change Memory Devices}
\label{sec:pcm-devices}
Phase-change memory exploits the behavior of certain chalcogenide materials that can be switched rapidly and repeatedly between amorphous and crystalline phases~\cite{Raoux2014}. These materials are typically compounds of Germanium, Antimony, and Tellurium (Ge\textsubscript{x}Sb\textsubscript{y}Te\textsubscript{z}, GST). The amorphous phase presents a high electrical resistivity while the crystalline phase exhibits low resistivity. A PCM device consists of a certain volume of this phase change material sandwiched between two electrodes (Fig.~\ref{fig:device}). Applying pulses to a PCM device results in Joule heating, which alters the phase (state) of the material. A reset pulse is used to melt a significant portion of the phase change material. When the pulse is stopped abruptly, the molten material quenches into the amorphous phase. Following the reset pulse, the device will be in a high resistive state (HRS). When a slower set pulse, with an amplitude above a threshold voltage ($V_{th}$)~\cite{Ovshinsky1968} is applied to a PCM device in the HRS, the amorphous region crystallizes. After the SET pulse, the device will be in a low resistive state (LRS). The resistance state can be read by biasing the device with a low read voltage that does not change the phase configuration.

Our setup (Fig.~\ref{fig:setup}) includes three PCM cells with a shared bottom electrode, and enables programming and reading each cell as well as performing the logic operations. A write-verify scheme is used to probe the maximum cycle count of a single device and characterize its switching behavior. For set and reset, we use 30/500/500~ns and 30/50/30~ns rise/width/fall pulses, respectively. The resistance is measured with a 0.2~V, $1~\mu s$ read pulse. We consider resistance higher than 100~k$\Omega$ as HRS (logical '0') and resistance lower than 10~k$\Omega$ as LRS (logical '1').
The current, voltage, and power required to set and reset the devices are depicted in Fig.~\ref{fig:characterize}a-c. Here, multiple pulses are used, while varying the voltage, until the set or reset event is encountered. Our endurance test shows that a device can maintain a 10X resistance window for almost 10\textsuperscript{4} cycles, with some degradation in the resistance distribution after several hundred cycles (Fig.~\ref{fig:characterize}d). The current and voltage waveforms across the PCM cell during a typical set operation are shown in Fig.~\ref{fig:characterize}e. The set operation is the basis for our proposed logic operations, where threshold switching occurs within $\sim$100ns. Finally, the current-voltage (I-V) transition of the device from the amorphous state to the crystalline state is shown in Fig.~\ref{fig:characterize}f.

\subsection{Phase-Change Memory Stateful Logic}
\begin{table}[t]
\centering
\setlength{\tabcolsep}{10pt}
\def\arraystretch{2.5}
\setlength{\arrayrulewidth}{0.56pt}
\caption{Summary of the applied voltages and configurations at each node to realize the logic gates}
\begin{threeparttable}
\begin{tabular}{c|c|c|c|c}
\hline
\hline
\multirow{2}{*}{\textbf{Gate}} & \multicolumn{4}{c}{\textbf{Voltage/Configuration}} \\
\cline{2-5}
& \textbf{TE\textsubscript{IN1}} & \textbf{TE\textsubscript{IN2}} & \textbf{TE\textsubscript{OUT}}* & \textbf{BE\textsubscript{ALL}}**\\
\hline
\hline
NOR & $\sim \frac{1}{2}V_{th}$ & $\sim \frac{1}{2}V_{th}$ & $\sim V_{th}$ & R \\
\hline
IMPLY & $\sim \frac{1}{2}V_{th}$ & F & $\sim V_{th}$ & R \\
\hline
OR & 0~V & 0~V & $\sim V_{th}$ & F \\
\hline
NIMP & $\sim V_{th}$ & $\sim \frac{1}{3}V_{th}$ & 0~V & F \\
\hline
\hline
\end{tabular}
\begin{tablenotes}
        
      \item * For the IMPLY gate, OUT serves also as input.
      \item ** 'R' marks connection to the grounded fixed resistor, 'F' marks a node left floating.
\end{tablenotes}
\end{threeparttable}
\label{tab:voltages-config}
\end{table}
The proposed PCM-based stateful logic gate consists of three PCM cells where two cells serve as inputs and the third device as the output (Fig.~\ref{fig:setup}). The output cell may also serve as an additional input at the cost of losing its stored data, $i.e$, a destructive operation. A grounded fixed resistor (10~k$\Omega$ in our configuration) can be connected to the shared node as well (similar to material implication in RRAM~\cite{Borghetti2010}). A logic operation is achieved by applying voltage pulses to the top electrode (TE) of the input cells, causing a conditional output switching, depending on the resistive states of the inputs. Note that the cells for IN1, IN2 and OUT are interchangeable, as memory cells in the same row.
The switching mechanism is based on the Ovonic threshold switching phenomenon~\cite{Ovshinsky1968} that occurs if the voltage across the output cell is above its threshold voltage, $V_{th}$, followed by the crystallization of the output cell. Since the logic operation is based on a switching event, the endurance data in Fig.~\ref{fig:characterize}d represents the degradation of the output device. The voltage selection and design methodology is similar to what is common in RRAM based stateful logic~\cite{Wald2016}, where voltage divider expressions are used to characterize the voltage distribution across the output and input cells.
Furthermore, this design principle is compatible with the crossbar memory structure commonly used for PCM~\cite{Ielmini2011,Sebastian2019}. Here, we propose four different logic functions (\textit{i.e.,} NOR, IMPLY, NOR, NIMP) based on this principle. Note that the fixed resistor required by some gates can be implemented as part of the peripheral circuitry for each word-line in the crossbar. Fig.~\ref{fig:crossbar} and Table~\ref{tab:voltages-config} summarize the voltages and configurations used to realize the different logic gates and their mapping to the crossbar structure. The crossbar structure is referred to as the general form of a crossbar memory with vertical bit-lines and horizontal word-lines. The same principles can be applied to state-of-the-art PCM crossbars~\cite{Gong2020} that mitigate the sneak-path and write-disturb phenomena by incorporating an Ovonic threshold selector (OTS) per cell, since the switching mechanism of an OTS is also based on a threshold voltage. However, in the presence of OTS, the distribution of voltages between the operating cells might change and affect the functionality of the gates. In future work, we plan to further examine and verify stateful logic using PCM-OTS crossbars.

\begin{figure*}[t]
\includegraphics[width=\linewidth]{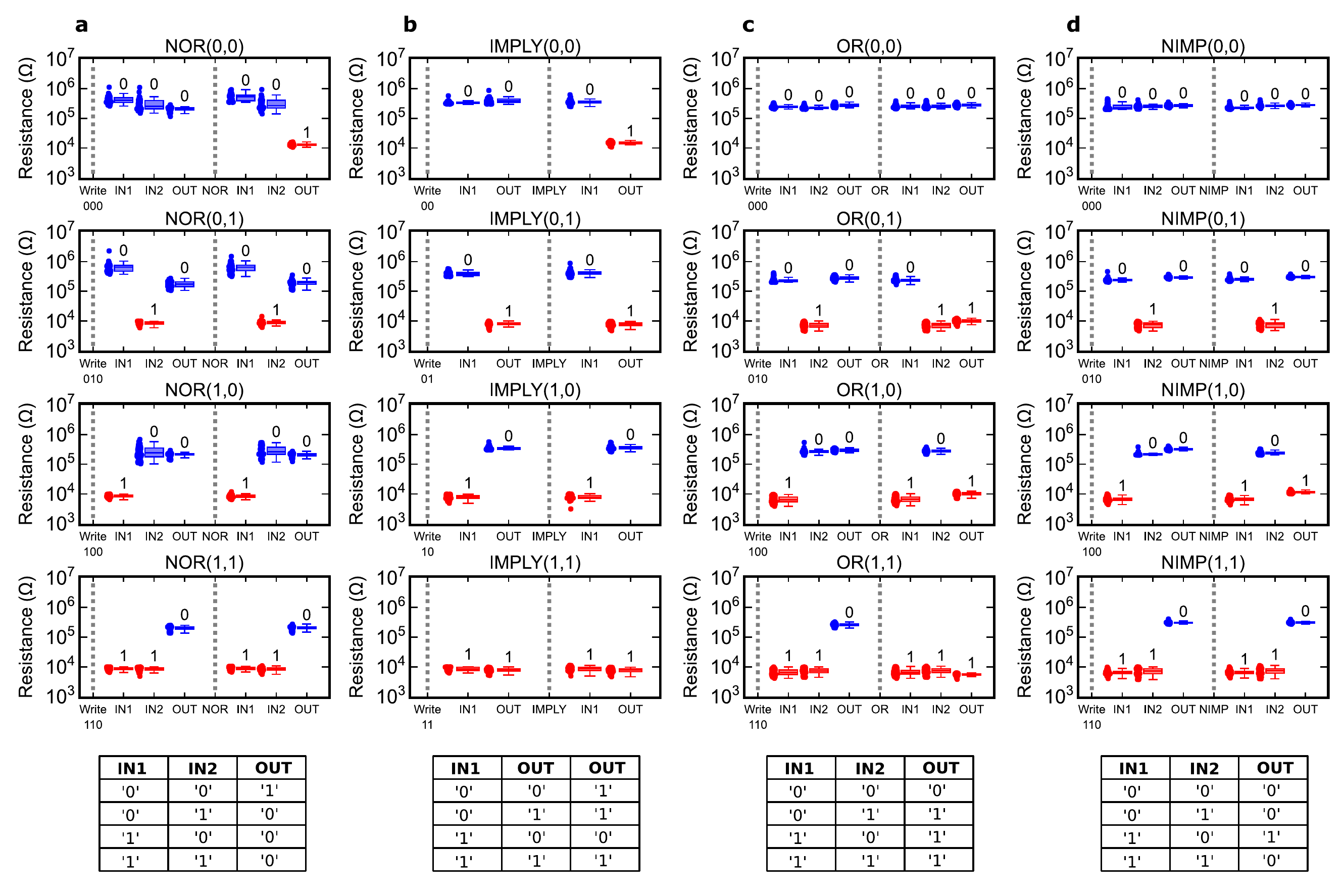}
\caption{Experimental results of the proposed logic gates. 50 iterations of the (a) NOR, (b) IMPLY, (c) OR, and (d) NIMP gates. The x-axis is the operation or read cell, the y-axis is the measured resistance plotted as a scatter and a median box. For the IMPLY gate, OUT is used also as input. All iterations of all the gates show correct logic operation and exhibit input stability.}
\label{fig:results}
\end{figure*}

A NOR operation is realized by initializing the output cell to the HRS and grounding the shared bottom electrode (BE) using the fixed resistor. Applying $\sim\frac{1}{2}V_{th}$ to TE\textsubscript{IN1},TE\textsubscript{IN2}, and applying $\sim V_{th}$ to TE\textsubscript{OUT} causes the voltage across OUT to be approximately $V_{th}$ only if both inputs are in HRS. Additionally, the state of the inputs remains unchanged because the maximum voltage across each input cell is $\sim~\frac{1}{2}V_{th}$. Similarly, a destructive implication (IMPLY) operation is realized. Here, TE\textsubscript{IN2} is kept floating, OUT serves as an input as well, and the voltage across OUT is $\sim V_{th}$ only if IN1 is in HRS. An OR operation is realized by keeping the shared BE floating, grounding TE\textsubscript{IN1}, TE\textsubscript{IN2}, and applying $V_{th}$ to TE\textsubscript{OUT}. This causes the voltage across the output cell to be approximately $V_{th}$ only if at least one input is in the LRS. Similarly, a not implication (NIMP) operation is realized by keeping the shared BE floating, grounding TE\textsubscript{OUT}, and applying $\sim V_{th}$ to TE\textsubscript{IN1} and $\sim \frac{1}{3}V_{th}$ to TE\textsubscript{IN2}. Here, the voltage across the output cell is approximately $V_{th}$ only if IN1 is in LRS and IN2 is in HRS. As described in previous works~\cite{Hoffer2020}, an XOR gate can be performed in two steps by running the NIMP operation twice on the same output with alternating inputs. Our logic set is functionally complete (NOR by itself is functionally complete), and synthesis tools can be used to determine the required execution steps of any desired logic function by applying sequential operations of these gates~\cite{BenHur2020}.

\begin{table}[!t]
\caption{Experimental Demonstrations of Stateful Logic}
\label{tab:stateful-comp}
\renewcommand{\arraystretch}{1.7}
\setlength{\tabcolsep}{7pt}
\setlength{\arrayrulewidth}{0.2mm}
\centering \small
\resizebox{\columnwidth}{!}{
    \begin{tabular}{P{0.5cm} P{2.0cm} P{2.0cm} P{2.0cm} P{2.0cm}}
        \hline
        \hline
         & \textbf{\textsf{Technology}} & \textbf{\textsf{Functions}} &
         \textbf{\textsf{No. Steps}} & \textbf{\textsf{No. Tests Reported}} \\
         \hline
         ~\cite{Borghetti2010} & RRAM & IMPLY & 1 & 1  \\
         \hline
         ~\cite{Adam2016} & RRAM & IMPLY & 1 & 80  \\
         \hline
         ~\cite{Jang2017} & RRAM & NOR, NOT & 1 & 50 \\
         \hline
         ~\cite{Bae2017} & RRAM & NOR, NOT & 1 & 1 \\
         \hline
         ~\cite{Yu2018} & RRAM & IMPLY & 1 & 1  \\
         \hline         
         ~\cite{Jang2018} & RRAM & NOR, NOT  & 1 & 1 \\
         \hline
         ~\cite{Zhong2018} & RRAM & NAND, NOR, NIMP, MAJORITY3, PARITY3 &  1 & 1 \\
         \hline
         ~\cite{He2019} & RRAM & IMPLY & 1 & 1  \\
         \hline         
         ~\cite{Kim2019} & RRAM & IMPLY, NIMP, AND, OR, NOR & 1 & 1 \\
         \hline
         ~\cite{Kim2020} & RRAM & NAND, NOR &  1 & 50 \\
         \hline
         ~\cite{Hoffer2020} & RRAM & OR, NIMP &  1 & 50 \\
         \hline
         ~\cite{Cassinerio2013} & PCM & NOT, AND & 2, 3 & 1 \\
         \hline
         \textbf{This Work} & PCM & NOR, IMPLY, OR, NIMP & 1 & 50 \\
         \hline
         \hline
    \end{tabular}
}
\vspace{-5pt}
\end{table}

\section{Experiments}
\subsection{Device Fabrication}
PCM devices are fabricated as in~\cite{Neumann2019}, starting with the evaporation and etching of tungsten (W) to form the bottom electrodes (BEs). Next, SiO\textsubscript{x} is deposited with plasma-enhanced chemical vapor deposition (PECVD) and pattern the confined vias using e-beam lithography. In-situ Ar sputtering is used to make sure the top of the BE is not oxidized. Then, sputtering and liftoff are used to pattern the Ge\textsubscript{2}Sb\textsubscript{2}Te\textsubscript{5} (GST) layer with $in~situ$ TiN capping, and the final TiN/Pt top electrodes (TEs) and contact pads. 

\subsection{Electrical Measurements}
The measurement setup is shown in Fig.~\ref{fig:setup}; it includes three PCM cells and enables programming and reading each cell, as well as performing the logic operations.
Electrical measurements are performed on-wafer using a Keysight B1500A with four B1530 WGFMU channels and a Keysight MSOX3104T oscilloscope. 

\subsection{Experimental Demonstration}
\begin{figure*}[t]
\vspace{-5pt}
\centering
\includegraphics[width=\linewidth]{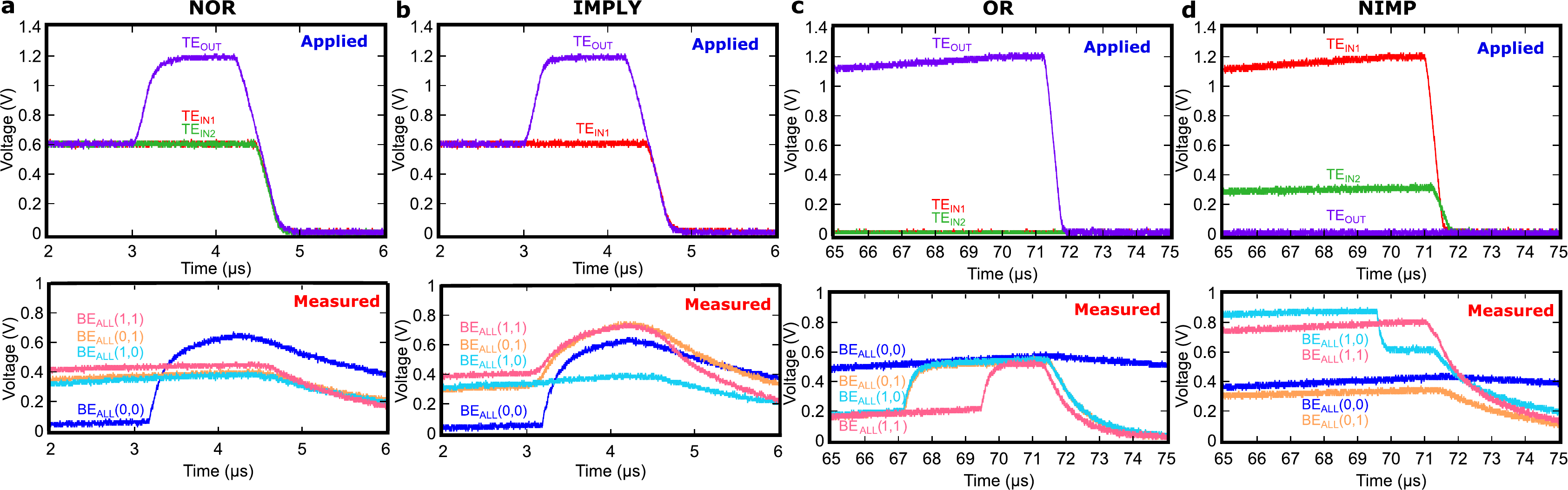}
\caption{Voltages applied to the TE of the input/output cells to evaluate the (a) NOR, (b) IMPLY, (c) OR, and (d) NIMP gates. The measured voltage on the BE shared node is plotted for each input case. (a) NOR gate; if at least one of the inputs is in LRS, the voltage at the shared node follows the constant voltage dictated by the inputs, thus keeping the voltage across OUT below the set threshold.
(b) IMPLY gate; the voltage at the shared node changes according to the resistive state of the inputs.
(c) OR gate; the voltage across OUT is higher than the set threshold if at least one input is in LRS, causing a switch event noticeable by the voltage change at the shared node.
(d) NIMP gate; the voltage across OUT is higher than the set threshold if IN1 is in LRS and IN2 is in HRS, causing a switch event noticeable by the voltage change at the shared node.}
\label{fig:waveforms}
\vspace{-5pt}
\end{figure*}
We measured the functionality and robustness of the proposed gates on the fabricated devices. In each test iteration, we examine the four input combinations for the tested gate. Each experiment includes: a) a write-verify procedure to initialize the inputs and output to the desired states, b) applying the voltage pulses required to evaluate the logic function, and c) cells reading to examine the output result and to verify the stability of the inputs. The write-verify procedure includes using the voltage pulses to set the cells to the desired state, verifying the result using a read pulse, and applying the voltage pulses again until the resistance is in the desired range. The voltage pulses and resistance ranges are as described in Section~\ref{sec:pcm-devices}. A stateful operation is evaluated not only by the correct logical result, but also the stability of the inputs. This is not always trivial, as reported by previous RRAM works~\cite{Siemon2019,Hoffer2020,Hyun2020,Kim2020}.
Results of 50 iterations for: (a) NOR, (b) IMPLY, (c) OR, and (d) NIMP logic gates are shown in Fig.~\ref{fig:results}. For each gate, all test iterations were performed sequentially on the same couple/triplet of devices. The results show successful logic operation for all iterations on all gates. Additionally, the inputs remain stable, without any meaningful change in their resistance. Therefore, the input degradation is negligible. However, we note that our work is a proof-of-concept and to have conclusive results regarding degradation, additional measurements on larger arrays are required. Table \ref{tab:stateful-comp} compares previous experimental demonstrations of stateful logic, primarily using RRAM, with our demonstration using PCM. Note that we do not compare energy and latency numbers, since the experimental demonstrations are proof-of-concepts and usually use unoptimized, university-fabricated devices that are hard to compare. Furthermore, the different measurement setup in each work may affect the results. Previous PCM stateful logic work requires 2-3 steps for computation, while the proposed PCM stateful logic uses a single step for the execution of different functions. We do not count output initialization as a computation step in this comparison since all methods require it. Compared to RRAM stateful logic, the complexity of our operations is similar, and the actual difference in terms of latency and energy lies on the different device properties and switching mechanisms. In future work, we plan to further examine the differences between RRAM and PCM based stateful logic methods using experimental measurements.

In the NOR gate test, we apply 0.6~V for 3~$\mu s$ on all TEs, then we increase the voltage on TE\textsubscript{OUT} to 1.2~V for 1~$\mu s$, while keeping the voltage on TE\textsubscript{IN1} and TE\textsubscript{IN2} at 0.6~V. We apply the same voltage scheme for the IMPLY gate test, but TE\textsubscript{IN2} is kept floating. The value for the fixed resistor in the NOR and IMPLY gates was selected as 10k$\Omega$, the max value for LRS, to assure the current passing through the output is high enough for complete crystallization during output switching, while keeping the voltage across the output lower than the threshold voltage for the non-switching cases.
In the OR gate test, we keep the shared BE floating, ground TE\textsubscript{IN1} and TE\textsubscript{IN2}, and apply 1.2~V on TE\textsubscript{OUT} with rise time of 70~$\mu s$ and pulse length of 1~$\mu s$. Similarly, for the NIMP gate we keep the shared BE floating, ground TE\textsubscript{OUT} and apply 1.2~V on TE\textsubscript{IN1} and 0.35~V on TE\textsubscript{IN2} with rise time of 70~$\mu s$ and pulse length of 1~$\mu s$.

The shape of the voltage pulses used for the demonstration of the NOR and IMPLY gates has two parts, whereas the pulses for the OR and NIMP gates have a relatively long rise time. These pulses were selected to compensate for the long RC delays to the shared node in our experimental setup, and will be substantially shorter in an integrated design. The RC delays are caused by a combination of the large parasitic capacitance of the pads and probes and the effective resistance that drives the shared node and differ between gates, due to their different circuits. In the OR and NIMP gates, this problem is most prominent in the IN1=IN2=`0' input case, since the impedance that drives the shared node is relatively high (the inputs and the output are in HRS). Without using a long rise time, the output might switch regardless of the state of the inputs or the inputs might change, since it takes time for the voltage at the shared node to update to its steady state voltage. For the NOR and IMPLY gates this issue is less distinct, since the circuit uses a small fixed resistor that connects the shared node to ground, which decreases the value of RC. Nevertheless, we chose using a pulse with two parts to deal with the RC delay for the NOR and IMPLY gates, since it is still meaningful. Although the RC delay here is significant since it is considerably large in the experimental setup, the results indicate that even in an integrated setup, parasitic capacitance might be a limiting factor that can affect crossbar size selection and the performance of the logic gates.

The applied voltage pulses to implement the gates and the measured voltage at the shared BE, marked as BE\textsubscript{ALL}, for each input case are depicted in~Fig.~\ref{fig:waveforms}. In the NOR and IMPLY tests, if one of the inputs is in LRS, the voltage at BE\textsubscript{ALL} follows the constant voltage dictated by the inputs, thus keeping the voltage across the output below the set threshold. Otherwise, the voltage at BE\textsubscript{ALL} remains at 0~V, and the output is switched once the voltage on its TE is above $V_{th}$. 
In the OR and NIMP tests, the voltage at BE\textsubscript{ALL} follows a constant trend for all non-switching cases, keeping the voltage across the output cell below $V_{th}$. In the cases where the output is switched, a meaningful change in the voltage at BE\textsubscript{ALL} is noticeable, caused by the resistance change of the output.
\vspace{-10pt}
\section{Conclusion}
To tackle the incessant data movement between the CPU and memory, we propose adding computation capabilities to PCM technology, inspired by previously proposed stateful logic for RRAM. Since the PCM switching mechanism is fundamentally different than RRAM, new circuits are required.
We experimentally demonstrate a new method to perform four stateful logic gates using PCM (NOR, IMPLY, OR, and NIMP) in a single step. The measured results show correct and robust logic operation with 50 test iterations demonstrated for each gate. The proposed gates are crossbar compatible, functionally complete and can be executed simultaneously on multiple rows. This may reignite scientific interest in the PCM technology, which was almost completely disregarded for stateful logic and paves the path towards PCM-based digital processing-in-memory architectures.

% \appendices
\vspace{-10pt}
\section*{Acknowledgement}
Fabrication was performed at the Stanford Nanofabrication Facility (SNF) and the Technion Micro-Nano Fabrication Unit (MNFU). 
\vspace{-15pt}
\bibliographystyle{IEEEtran2.bst}
\bibliography{IEEEabrv,./ref}

\end{document}